\documentclass[aps,prc,reprint,groupedaddress,twocolumn,superscriptaddress]{revtex4-2}

\usepackage{amssymb}
\usepackage{amsmath}
\usepackage{graphicx}
\usepackage{color}
\usepackage{braket}
\usepackage{siunitx}

\begin{document}


\title{Direct demonstration of the two-phonon structure of the $J^{\pi} = 1_{\qty{4742}{keV}}^{-}$ state of $^{88}$Sr}


\author{J.~Isaak}
\email{jisaak@ikp.tu-darmstadt.de}

\affiliation{Technische Universit\"at Darmstadt, Department of Physics, Institute for Nuclear Physics,
            64289 Darmstadt, Germany}

\author{D.~Savran}
\affiliation{GSI Helmholtzzentrum für Schwerionenforschung GmbH, 64291 Darmstadt, Germany}

\author{N.~Pietralla}
\affiliation{Technische Universit\"at Darmstadt, Department of Physics, Institute for Nuclear Physics,
            64289 Darmstadt, Germany}

\author{N.~Tsoneva}
\email{nadia.tsoneva@eli-np.ro}
\affiliation{Extreme Light Infrastructure (ELI-NP), Horia Hulubei National Institute for R \& D in Physics and Nuclear Engineering (IFIN-HH), 077125 Bucharest-Magurele, Romania}
\affiliation{Institut für Theoretische Physik, Universit\"at Gießen, Heinrich-Buff-Ring 16, D-35392 Gießen, Germany}
             
\author{A.~Zilges}
\affiliation{Institut f\"ur Kernphysik, Universit\"at zu K\"oln, 
             50937 K\"oln, Germany}

\author{K.~Eberhardt}
\affiliation{Forschungsreaktor TRIGA Mainz, Johannes Gutenberg-Universit\"at Mainz, 55128 Mainz, Germany}

\author{C.~Geppert}
\affiliation{Forschungsreaktor TRIGA Mainz, Johannes Gutenberg-Universit\"at Mainz, 55128 Mainz, Germany}

\author{C.~Gorges}
\affiliation{Forschungsreaktor TRIGA Mainz, Johannes Gutenberg-Universit\"at Mainz, 55128 Mainz, Germany}

\author{H.~Lenske}
\affiliation{Institut für Theoretische Physik, Universit\"at Gießen, Heinrich-Buff-Ring 16, D-35392 Gießen, Germany}

\author{D.~Renisch}
\affiliation{Department Chemie - Standort TRIGA, Johannes Gutenberg-Universit\"at Mainz, 55128 Mainz, Germany}
\affiliation{Helmholtz Institut Mainz, 55128 Mainz, Germany}



\date{\today}

\begin{abstract}


We have studied the decay pattern of the $J^{\pi} = 1_{\qty{4742}{keV}}^{-}$ state of $^{88}$Sr to probe its 
quadrupole-octupole coupled two-phonon structure. 
In particular, a unique fingerprint to prove the 
two-phonon nature is the $E2$ decay strength of the 
$1_{\qty{4742}{keV}}^{-} \to 3_{1}^{-}$ transition 
into the one-octupole-phonon state. 
$\gamma$-ray spectroscopy was performed after the $\beta$-decay of $^{88}$Rb 
to obtain the necessary sensitivity for this weak-intensity decay branch. 
Sufficient amounts of $^{88}$Rb (T$_{1/2}$ = 17.8 min) were produced by neutron activation of natural Rb in the TRIGA Mark II reactor. 
The results show that the $B(E2)$ value of the $1_{\qty{4742}{keV}}^{-} \to 3_{1}^{-}$ transition is equal to the $B(E2)$ of the 
$2_{1}^{+} \to 0_{1}^{+}$ transition, directly demonstrating the 
quadrupole-octupole coupled two-phonon nature of the $1^-$ state. 
A comparison of the results with energy-density functional plus quasiparticle-phonon model calculations shows remarkable agreement, corroborating this assignment.

\end{abstract}


\maketitle


\section{Introduction}

Low-lying collective excitations of vibrational nuclei near
shell closures are of particular interest for nuclear
physics because atomic nuclei represent prime examples of strongly-coupled 
mesoscopic many-body quantum systems where collective vibrational 
excitations~\cite{bohr52}, frequently addressed as {\it phonons}, 
compete with single-particle excitations at low energy. 
Since the 1950s, there has been an intriguing question of how 
phonon excitations can couple to multi-phonon structures and thereby 
serve as building blocks for nuclear structure 
(see, e.g., ~\cite{cast90} and references therein). 
Prime examples in even-even nuclei are the low-lying $2^{+}$ and $3^{-}$ states that are described by quadrupole and octupole phonon oscillations of the nuclear surface, respectively.
The nature of a one-phonon state is described by the coherent superposition of particle-hole (p-h) excitations built on the ground state,
while two-phonon states are composed of 2p-2h excitations built on the ground state with a similar generalization for multi-phonon states. 
In a simple phonon-coupling picture, phonons can couple to multi-phonon quantum states forming 
multiplets with the same structure and correlated decay properties. 
In a purely harmonic oscillator, two phonon states occur at an excitation energy corresponding to the sum energy of the constituent phonons. Deviations from that energy estimate are addressed as anharmonicities.
Coupling of identical phonons based on the same single-particle excitations, e.\,g. $(2^{+} \otimes 2^{+})$ \cite{scha55, kern95, garr18} or $(3^{-} \otimes 3^{-})$ \cite{yeh96, klei82}, 
may lead to strong anharmonicities due to Pauli blocking effects.
In contrast, Pauli blocking can be partly avoided by coupling different phonons, i.\,e., inhomogeneous phonon coupling~\cite{piet99} when the constituent phonons exhibit different single-particle structures.
In leading order, the distortions are expected to be small in such inhomogeneous multi-phonon states and, thus, allow for a stringent test of the simple multi-phonon picture of collective low-lying excitations, both, in terms of anharmonicities and of purity of the multi-phonon wave functions.
One prominent example of such an inhomogeneous phonon coupling is the quadrupole-octupole two-phonons excitations ($2^{+} \otimes 3^{-}$) forming a quintuplet of nuclear levels with spin-parity quantum numbers $J^{\pi} = 1^{-}, 2^{-}, 3^{-}, 4^{-}$, and $5^{-}$ 
\cite{lipa66, voge71, smir2000}.

The criteria for the solid experimental identification of vibrational multi-phonon states 
are currently still under debate~\cite{garr18, garr20}.
Lacking other experimental information, the excitation energy of the multi-phonon candidate 
states has often been considered for a multi-phonon criterion because it is expected to be 
close to the sum energy of the constituent phonons when assuming their harmonic coupling. 
This criterion is, however, weak because anharmonicities lead to deviations from this expectation 
and nuclear levels with different structure that, for any reason, may coincide with 
the sum energy, can erroneously be misinterpreted as multi-phonon states although they are not. 
Instead, the decisive signatures for multi-phonon structures are the collective  strengths 
of the corresponding phonon-annihilating decay transitions that should correlate to the 
strengths of the ground-state decays of the corresponding one-phonon states. 
Anharmonicities observed for multi-phonon states may be caused by the influence 
of the underlying single-particle degree of freedom. 
Hence, the clear establishment of multi-phonon states and the thorough study of their 
properties provide information on the interplay of the collective and single-particle 
degrees of freedom in atomic nuclei.

The scope of the present work is focused on the investigation of the structure of the candidate for the quadrupole-octupole coupled 
$J^{\pi} = 1^{-}_{2ph}$ state of $^{88}$Sr. 
Following the above mentioned signatures of a two-phonon state, the excitation energy is expected at $E(1^{-}_{2ph}) \approx E(2^{+}_{1}) + E(3^{-}_{1}) = \qty{1836}{keV} + \qty{2734}{keV} = \qty{4570}{keV}$. 
The method of Nuclear Resonance Fluorescence (NRF) has proven a powerful tool for 
searching for $1_{2ph}^{-}$ states of vibrational 
nuclei \cite{metz76, metz78, metz78b, herz95, knei96, piet99, andr01, fran04, zilg22}.
Previous NRF studies have identified the $1^{-}$ state at \qty{4742}{keV} to be the best candidate for the $1^{-}_{2ph}$ state of this nucleus~\cite{piet02c, piet02cerr, kaeu04}.
However, for a decisive identification, the reduced decay strengths of a harmonically coupled two-phonon $1^{-}_{2ph}$ state have to fulfill the following relations:
\begin{align}
\label{eq::decaystrength}
B(E3, 1^{-}_{2ph} \to 2^{+}_{1}) = B(E3, 3^{-}_{1} \to 0^{+}_{1}) \\
\label{eq::decaystrengthb}
B(E2, 1^{-}_{2ph} \to 3^{-}_{1}) = B(E2, 2^{+}_{1} \to 0^{+}_{1}) 
\end{align} 
It is experimentally challenging to determine the $E3$ contribution in the $1^{-}_{2ph} \to 2^{+}_{1}$ decay, due to the dominant competition by $E1$ multipolarity. 
The situation is different for the $1^{-}_{2ph} \to 3^{-}_{1}$ transition, since 
the $E2$ decay associated with the corresponding annihilation of the constituent quadrupole phonon 
is the leading multipolarity and a potential $M3$ multipole admixture is subdominant.
Using the value for $B(E2, 2^{+}_{1} \to 0^{+}_{1}) = \qty{176(9)}{e^2 fm^4}$ (adopted from~\cite{mccu14}) and the ground-state transition strength of the $1^{-}_{2ph}$ state (adopted from~\cite{kaeu04}),
one expects a branching ratio of $\Gamma(1^{-}_{2ph} \to 3^{-}_{1}) / \Gamma(1^{-}_{2ph} \to 0^{+}_{1}) = 0.034(5)$. 
The measurement of branching ratios in the regime of a few percent is challenging in NRF experiments 
due to the rapidly increasing low-energy background originating from nonresonant photon scattering 
off the target. 
Some of us performed NRF experiments on $^{88}$Sr with quasi-monochromatic photon beams, but were unable to observe the 2008-keV $\gamma$-ray transition from the $1^{-}_{2ph}$ state to the $3^{-}_{1}$ state above background~\cite{piet02c, piet02cerr}.

An alternative approach to photon scattering experiments for the study of decay properties of excited states is off-beam $\gamma$-ray spectroscopy following $\beta$-decay after neutron activation. 
In contrast to in-beam experiments, activity measurements have the advantage of a significantly suppressed low-energy $\gamma$-ray background enabling the observation of weak transitions.
Therefore, the neutron activation of $^{87}$Rb was exploited to populate excited states of $^{88}$Sr via $\beta$ decay from $^{88}$Rb. 
The corresponding $Q(\beta^{-}) = \qty{5312}{keV}$ is sufficiently high to allow the population of the candidate $1^{-}_{2ph}$ two-phonon state at \qty{4742}{keV} and study its decay behavior.
Indeed the ground-state decay of the \qty{4742}{keV} $1^{-}_{2ph}$ state was observed in 
the $\beta$-decay of $^{87}$Rb in previous experiments~\cite{bunt76, miya02}.
However, other decay channels were not observed due to the limited counting statistics of previous measurements.
Therefore, we performed a new experiment with much higher sensitivity at the TRIGA Mark II reactor at Johannes Gutenberg-Universit\"at Mainz.

This Letter is organized as follows. In the next section, the details on the neutron-activation measurement at TRIGA Mark II are given and the analysis of the activated $^{87}$Rb sample is presented. We then discuss the results for $^{88}$Sr in comparison to other nuclei before the Letter is concluded.

\section{Experiment \& Analysis}
\label{sec:exp}
The neutron-activation experiment was conducted at the TRIGA Mark II
research reactor at Johannes Gutenberg-Universit\"at Mainz~\cite{eber07, eber19}. The TRIGA Mainz is a light water cooled
reactor using an alloy of uranium, zirconium, and hydrogen ($U_{0.03} Zr_{1} H_{1}$) as fuel material. The uranium is enriched to below \qty{20}{\percent} in $^{235}$U. At the chosen irradiation position a neutron flux of about $10^{12}$~neutrons per cm$^{2}$s is achieved at \qty{100}{kW} reactor power.

Four samples with \qty{500}{ml} of RbNO$_3$ solution with a concentration of
\qty{100}{mg/ml} in natural Rb were irradiated for \qty{30}{min} each, resulting in
an activity of about \qty{5.7}{MBq} of $^{88}$Rb, each. 
After a short waiting and
transport time, the samples were placed in front of a high-purity
germanium (HPGe) detector with a photo-peak efficiency at \qty{1.33}{MeV} of \qty{90}{\percent} relative to an $\qty{3}{''} \times \qty{3}{''}$ NaI detector. The detector is 
housed in a massive
lead and copper shielding in order to reduce contributions of natural
background. For the counting measurement, the samples were placed in an aluminum
cylinder with a wall thickness of \qty{9.13}{mm} in order to stop the
electrons emitted in the beta decays. 
The setup allows to change the
detector-to-sample distances in steps of \qty{2}{cm} up to a maximum distance
of \qty{24}{cm}. The time for a single counting measurement was set to \qty{5}{min}. Due to the short half-life of
$^{88}$Rb of about \qty{17.8}{min}, the overall count rate in the HPGe
detector drops rapidly within a few minutes. In order to compensate for the reduced activity, the
target was moved closer to the detector in steps of \qty{2}{cm} when at the end
of a counting measurement run the count rate dropped below \qty{3}{kcps} (kilo counts per second). The closest detector-to-sample distance
used was \qty{6}{cm} to avoid considerable summing effects, which would become significant at even shorter distances. 
Summing effects are taken into account in the analysis using an iterative procedure with \texttt{GEANT} simulations, discussed in more detail below.

\begin{figure}
  \includegraphics[width=0.9\columnwidth]{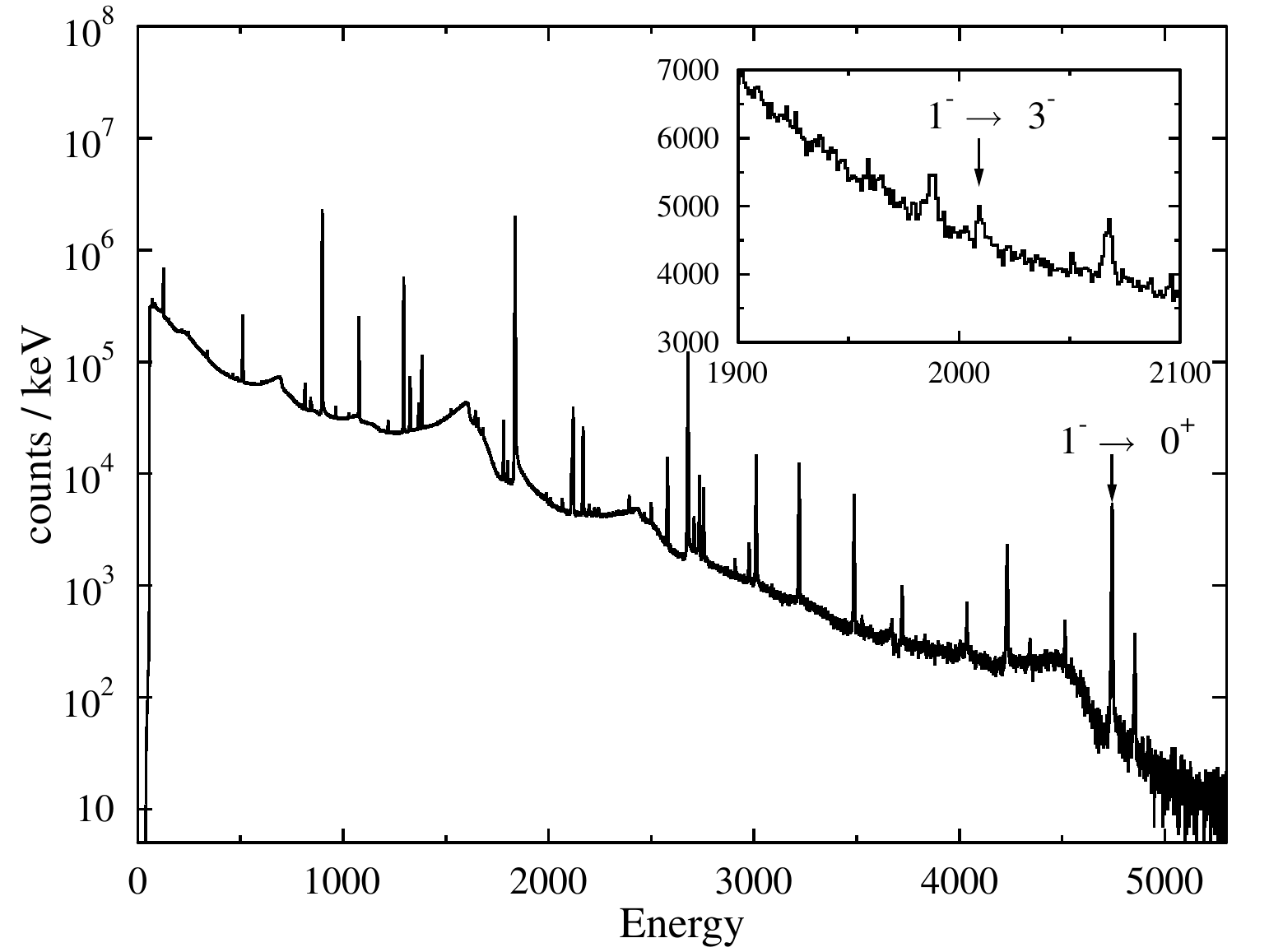}%
\caption{Summed $\gamma$-ray spectrum of all four neutron-activation measurements taken with the HPGe detector. The ground-state decay of the candidate for the $1^{-}_{2ph}$ state is marked with an arrow. The inset shows the region of interest for the $1^{-}_{2ph} \rightarrow 3_1^-$ transition.\label{fig::spec}}
\end{figure}

\begin{figure}
  \includegraphics[width=0.9\columnwidth]{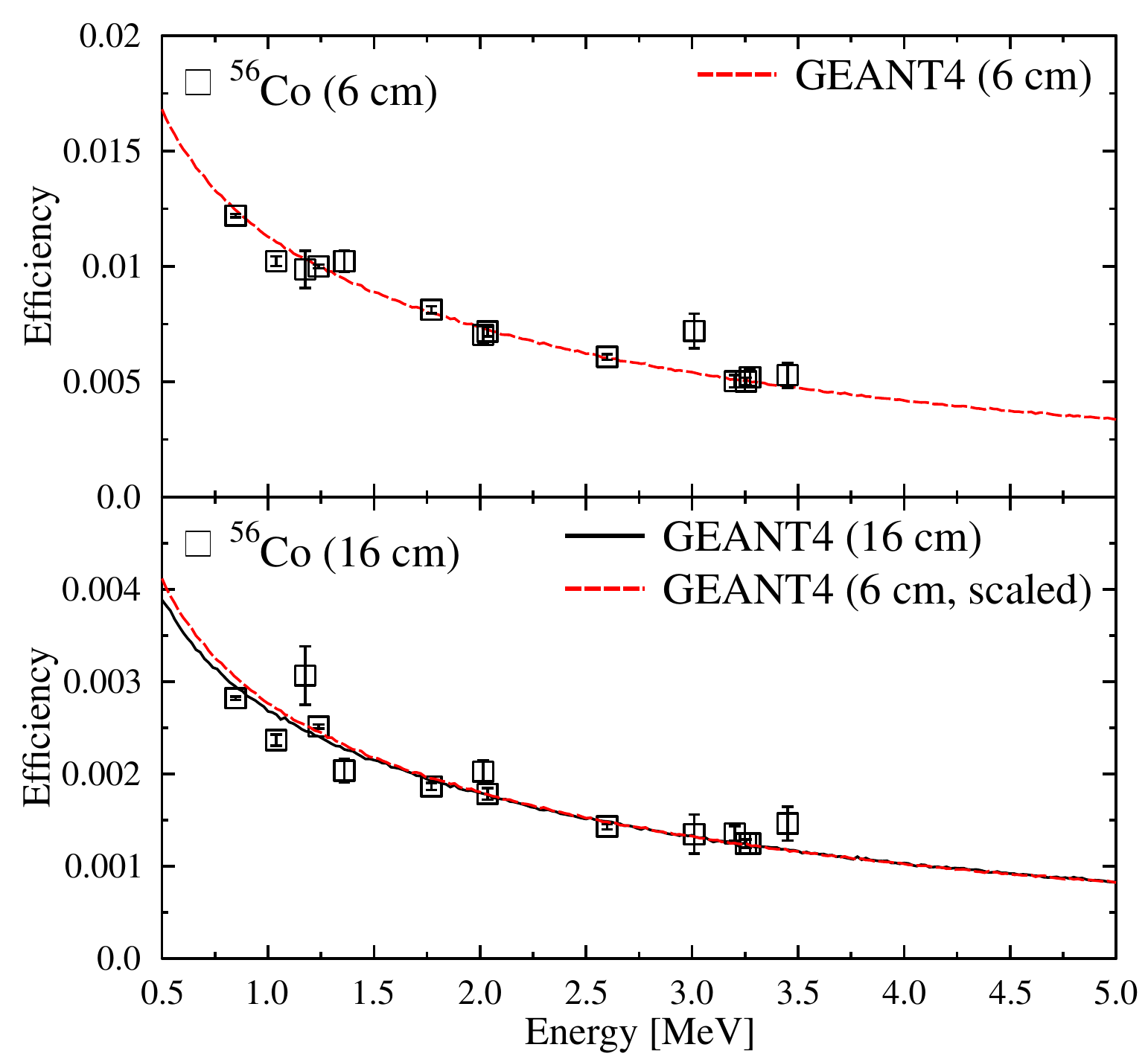} \caption{Simulated
photo-peak efficiencies for detector-to-sample distances of \qty{6}{cm} (upper panel) and \qty{16}{cm} (lower panel), respectively. 
The values measured with $^{56}$Co are scaled to
the simulations. For comparison of the energy dependence at different
distances the result of the simulation for \qty{6}{cm} is scaled and plotted
in the lower panel.\label{fig::eff}}
\end{figure}

Figure~\ref{fig::spec} displays the summed $\gamma$-ray spectrum
obtained for all neutron-activation runs taken with the four samples. The dominant
peaks belong to the beta decay of $^{88}$Rb, only small
contributions of $^{86}$Rb, $^{24}$Na and natural background are
observed. 
The decay of the $1^{-}$ state of interest at \qty{4742.5}{keV} to
the ground state is the dominant peak at higher energies. In total
29604(175) counts were collected in this transition, resulting in a
statistical uncertainty of about \qty{0.6}{\percent}. 
The inset of Fig. \ref{fig::spec} depicts the energy region of the expected $1^{-}_{2ph} \rightarrow 3_{1}^{-}$
transition at \qty{2008.3}{keV}, which is clearly visible above background
with 2246(49) counts in total. Also the decay into the $2_{1}^{+}$ 
and the $2_{2}^{+}$ states is observed, which was known already from~\cite{kaeu04}.
The branching ratio 
\begin{align}\label{eq::branching}
b_{f} = \frac{A_{1^{-} \rightarrow f} / \epsilon(E_{1^{-}} -
E_{f})}{A_{1^{-} \rightarrow 0_{1}^{+}} / \epsilon(E_{1^{-}})} 
\end{align} 
represents the decay intensity from the $1^{-}_{2ph}$ state to a state $f$ at excitation energy $E_{f}$ relative to its ground-state decay.
The corresponding peak areas determined from the $\gamma$-ray spectrum given in Fig.~\ref{fig::spec} are $A_{1^{-} \rightarrow f}$ and $A_{1^{-} \rightarrow
0_{1}^{+}}$, while the photo-peak detection efficiencies are denoted as $\epsilon(E_{1^{-}}-E_{f})$ and $\epsilon(E_{1^{-}})$, respectively.
The efficiency ratio in Eq.~(\ref{eq::branching}) is obtained from a \texttt{GEANT4} simulation~\cite{agos03, alli06, alli16} of an isotropically emitting $\gamma$-ray source placed in the experimental setup at detector-to-sample distances corresponding to the ones used in the actual measurements.
In order to check the accuracy of the simulation, measurements with a $^{56}$Co source were conducted at \qty{6}{cm} and \qty{16}{cm}, respectively.
The $^{56}$Co source was placed
in the same aluminum cylinder as the Rb samples. The result of the
simulation of the photo-peak efficiency for the two corresponding
distances of \qty{6}{cm} and \qty{16}{cm} is shown in Fig. \ref{fig::eff}. 
The experimental data points extracted
from the $^{56}$Co measurement are scaled to the simulation, because
the activity of the source is not known with sufficient
accuracy. Since the determination of the 
branching ratios $b_{f}$ relies on the
relative efficiencies only, it is sufficient to benchmark the
energy dependence of the simulated photo-peak
efficiency against the $^{56}$Co measurements. The comparison of the results 
shows that the variation of the energy dependence is weak for the two distances, especially for $\gamma$-ray energies above \qty{1.5}{MeV}, which is
the relevant energy region for the present analysis.
To quantify the accuracy of the simulations with respect to the scaled $^{56}$Co data points, the weighted
(external) uncertainty of the fit is calculated, which accounts for
the spread of the data about the fit. For the two distances of \qty{6}{cm} and \qty{16}{cm}, relative weighted uncertainties of
\qty{0.75}{\percent} and \qty{1.37}{\percent} are determined, respectively, which illustrates the good agreement of
the simulated energy dependence of the photo-peak efficiency to the experimental
values.

At the closest distance of \qty{6}{cm}, the total detection probability,
i.e. that any amount of energy is deposited in the detector, for a \qty{2}{MeV} $\gamma$-ray is about \qty{3.7}{\percent}. 
Thus, summing effects of
$\gamma$-rays within the cascades of the decay of the $1^{-}_{2ph}$ state
can play a role on that level. 
In order to account for this, the decay of
the $1^{-}_{2ph}$ state including the cascade transitions are simulated
using the same \texttt{GEANT4} simulation. The resulting intensities are
compared to the experimental results in an iterative procedure: In the
first step the branching ratios determined from experiment using
Eq. (\ref{eq::branching}) are used as input for the simulation of the
cascades. The intensities obtained from this simulation are then
compared to the experimental ones and a new set of branching ratios are
determined to be used in the next iteration step. This procedure
quickly converges to a final set of branching ratios for the different
decay paths of the $1^{-}_{2ph}$ state of interest.  

Repeating this analysis using simulations at different distances
results in slightly different final branching ratios, since the effect
of summing depends on the total detection efficiency. Thus, the
different results have been averaged weighted with the amount of data
taken at the corresponding distance.

\section{Results \& Discussion}
\label{sec:disc}

The experimentally determined branching ratios $b_f$ are summarized in Table~\ref{tab::results}. 
With the present experiment, the uncertainties of the decay branching ratios to the $2^{+}_{1,2}$ states and to the $3^{-}_{1}$ state are improved by one order of magnitude enabling a determination of the
ground-state branching ratio $\Gamma_{0}/\Gamma = 0.811(5)$ of the $1^{-}_{2ph}$ state at \qty{4742.4}{keV} to a precision of \qty{0.6}{\percent}, which is one order of magnitude smaller compared to the precision of about \qty{7}{\percent} reported in Ref.~\cite{kaeu04}.
Using this precisely determined value for $\Gamma_{0}/\Gamma$ and the literature value for $\Gamma_{0}^{2}/\Gamma = \qty{107(14)}{meV}$ from NRF experiments with bremsstrahlung~\cite{kaeu04}, one obtains $B(E1, 1^{-}_{2ph} \to 0^{+}_{1}) = \qty{1.18(16)}{\times 10^{-3} e^2 fm^2}$ for the reduced $E1$ transition strength to the ground state. See Table~\ref{tab::results_1minus} for a summary of the properties of the $1^{-}_{2ph}$ state.
\setlength{\tabcolsep}{9pt}
\renewcommand{\arraystretch}{1.25}
\begin{table}[]
    \centering
    \caption{Branching ratios $b_f$ for the observed decay channels of the $1^{-}_{2ph}$ state of $^{88}$Sr at \qty{4742.4}{keV} to the $2^{+}_{1}$, $3^{-}_{1}$, and $2^{+}_{2}$ states extracted from the present data. The $\gamma$-ray transition energy, the excitation energy of the final state and its spin-parity quantum number are given by $E_{\gamma}$, $E_{f}$, and $J^{\pi}_{f}$, respectively.}
    \begin{tabular*}{\columnwidth}{ccccc}
        \hline
        $E_\gamma$ [keV] & $E_f$ [keV] & $J_f^\pi$ & $b_{f}$ [\%] & $b_{f}$\footnote{computed from $\Gamma_{1,2}/\Gamma$ values given in Table~II of Ref.~\cite{kaeu04}} [\%] \\
        \hline
        4742.4      & 0         & $0^{+}_{1}$   &  100.0(6)       & 100.0 \\
        2906.3	    & 1836.1    & $2^{+}_{1}$   &  5.00(26)  & 4.4(19) \\
        2008.3	    & 2734.1    & $3^{-}_{1}$   &  3.72(29)  & - \\
        1523.9	    & 3218.5    & $2^{+}_{2}$   &  14.5(6) & 22(9) \\
        \hline
    \end{tabular*}
    \label{tab::results}
\end{table}

Moreover, the direct transition from the $1^{-}_{2ph}$ two-phonon candidate state to the $3^{-}_{1}$ 
one-octupole phonon state at \qty{2734.1}{keV} was observed in the present experiment with a branching ratio of $b_{f} = \qty{3.72(29)}{\percent}$ for the first time. 
Its measurement provides now the possibility to determine the corresponding $E2$ decay rate. 
\setlength{\tabcolsep}{5.5pt}
\renewcommand{\arraystretch}{1.25}
\begin{table}[]
    \centering
    \caption{Properties of the $1^{-}_{2ph}$ state at \qty{4742.4}{keV}. The ground-state branching ratio $\Gamma_{0}/\Gamma$, ground-state transition width $\Gamma_{0}$, total width $\Gamma$, and the reduced ground-state transition strength $B(E1, 1^{-}_{2ph} \to 0^{+}_{1})$ are determined using the branching ratios given in Table~\ref{tab::results}.}
    \begin{tabular*}{\columnwidth}{ccccc}
        \hline
        $\Gamma_{0}^{2}/\Gamma$\footnote{Value computed from $g \Gamma_{0}^{2}/\Gamma = \qty{322(42)}{meV}$ given in Table I of Ref.~\cite{kaeu04} using $g = 3$.} & $\Gamma_{0}/\Gamma$ & $\Gamma_{0}$ & $\Gamma$ & $B(E1, 1^{-} \to 0^{+}_{1})$ \\
        \text{[meV]} & & [meV] & [meV] & [$10^{-3}$e$^2$fm$^2$] \\
        \hline
        107(14) & 0.811(5) & 132(17) & 163(21) & 1.18(16) \\
        \hline
    \end{tabular*}
    \label{tab::results_1minus}
\end{table}
The resulting reduced $E2$ transition strength amounts to $B(E2, 1^{-}_{2ph} \to 3^{-}_{1}) = \qty{186(28)}{e^2 fm^4}$. 
As discussed above, a strong $E2$ transition from the $1^{-}_{2ph}$ state to the $3^{-}_{1}$ is expected.
In the two-phonon interpretation, the $E2$ decay strength is supposed to be equal to the quadrupole-phonon annihilation depopulating the $2^{+}_{1}$ state.
A comparison to the adopted value for $B(E2, 2^{+}_{1} \to 0^{+}_{1}) = \qty{176(9)}{e^2 fm^4}$ from Ref.~\cite{mccu14} yields a ratio of:
\begin{align}\label{eq::be2ratio}
\frac{B(E2, 1^{-} \to 3^{-}_{1})}{B(E2, 2^{+}_{1} \to 0^{+}_{1})} = 1.06(17)
\end{align} 
This result is in excellent agreement with the interpretation of the $1^{-}_{2ph}$ state as a quadrupole-octupole coupled two-phonon state. 

The experimental observations agree with predictions of the energy-density functional (EDF) and quasiparticle-phonon model (QPM)~\cite{Tso16}. 
The building blocks of the QPM model basis are the quasiparticle-random-phase approximation (QRPA) phonons, which are superpositions of two-quasiparticle creation and annihilation operators ~\cite{solo92}. The wave function of an excited QPM state is treated as a superposition of one-, two-, and three-phonon components, which in this case result from the coupling of $J^{\pi}$=1${^{\pm}}$-7${^{\pm}}$ QRPA phonons and excitation energies up to the experimental endpoint excitation energy of the residual nucleus. For the dipole excitations, one-phonon states up to E$_{x}$=25 MeV are additionally considered, so that the pygmy dipole resonance (PDR) and the isovector giant dipole resonance (GDR) contribution to the E1 transitions of the low-lying  1$^ {- }$ states are explicitly taken into account avoiding the consideration of  effective charges. Multi-phonon components in the wave function lead to violation of the Pauli principle. To account for it properly, we used the exact commutation relations between the phonon creation and annihilation operators.

QPM transition operators have harmonic and anharmonic parts and can therefore consider transitions beyond the quasi-boson approximation. Namely, the harmonic part gives the contribution arising from transitions between components in the wave function that differ by one phonon, and the anharmonic part couples components each with the same number of phonons or between those that differ by an even number of phonons \cite{pono98}. In case of decay of the first 1$^-$ state to the ground state, the later one couples the two-phonon components of the first 1$^-$ state to the ground state and thus allows the description of this E1 transition.

In the QPM calculations, two $1^{-}$ states are obtained below 6~MeV. The wave function of the calculated $1^{-}_{1}$ state contains about 90~\% of the ($2^{+}_{1} \otimes 3^{-}_{1}$) component and 7 ~\% of the $1^{-}_{3}$ QRPA state, which is an almost pure neutron QRPA state associated with the excitation of the PDR mode in this nucleus ~\cite{schw08,schw13}. 
Furthermore, due to the one-phonon and two-phonon components in the structure of the $1^{-}_{1}$ state, anharmonicity effects shift the excitation energy of the state to lower excitation energies with respect to the pure harmonic expectation energy resulting from the sum of the energies of the 
QRPA $2^{+}_{1}$ and $3^{-}_{1}$ states. 

Thus, the theoretical results support the picture of a dominant contribution of quadrupole and octupole phonons in the wave function of the $1^{-}_{1}$ state with equal $B(E2)$ values in the $1^{-} _ {1} \to 3^{-}_{1}$ and $2^{+}_{1} \to 0^{+}_{1}$ transitions, interpreted by the annihilation of the same quadrupole phonon, represented here by the $2^{+}_{1}$ state with an $E2$ transition strength of about \qty{8}{W{.} u{.}} to the  ground state.
The $1^{-}_{2}$ is dominated to 98~\% by the ($2^{+}_{2} \otimes 3^{-}_{1}$) component. 
It can be ruled out as a potential representation of the experimentally observed $1_{2ph}^{-}$ state due to its calculated $B(E2, 1^{-}_{2} \to 3^{-}_{1}) = \qty{0.03}{W{.} u{.}}$ which exhibits a two orders of magnitude smaller $E2$ transition strength.

Theoretically, we identify the $1^{-}_{1}$ state in the QPM calculations as the quadrupole-octupole coupled two-phonon state of interest for comparison with the experimental data.
The computed transition strengths for the $1^{-}_{1}$ and $2^{+}_{1}$ states are in remarkable agreement to the experimental values and are summarized in Tab.~\ref{tab::qpm}.
The decay strengths for both the $1^{-}_{1} \to 0^{+}_{1}$ and the $1^{-}_{1} \to 3^{-}_{1}$ transitions are in excellent agreement to the measured data. 
\setlength{\tabcolsep}{9pt}
\renewcommand{\arraystretch}{1.25}
\begin{table}[]
    \centering
    \caption{Comparison of experimental results to theoretical calculations within the QPM for $^{88}$Sr.}
    \begin{tabular*}{\columnwidth}{lcc}
        \hline
         & Experiment & QPM  \\
        \hline
        $E_x(1_{1}^{-})$ [keV] & \qty{4742}{} & \qty{4603}{} \\
        $E_x(2_{1}^{+})$ [keV] & \qty{1836}{} & \qty{1830}{} \\
        $E_x(3_{1}^{-})$ [keV] & \qty{2734}{} & \qty{2760}{} \\
        \hline
        $B(E1, 1^{-}_{1} \to 0^{+}_{1})$ [mW.u.]   & $\qty{0.93(12)}{}$      & $\qty{0.96}{}$  \\
        $B(E2, 1^{-}_{1} \to 3^{-}_{1})$  [W.u.]  & $\qty{8.0(12)}{}$    & $\qty{8.1}{}$   \\
        $B(E2, 2^{+}_{1} \to 0^{+}_{1})$ [W.u.]   & $\qty{7.6(4)}{}$    & $\qty{8.0}{}$   \\
        \hline
    \end{tabular*}
    \label{tab::qpm}
\end{table}

\begin{figure}
  \includegraphics[width=0.85\columnwidth]{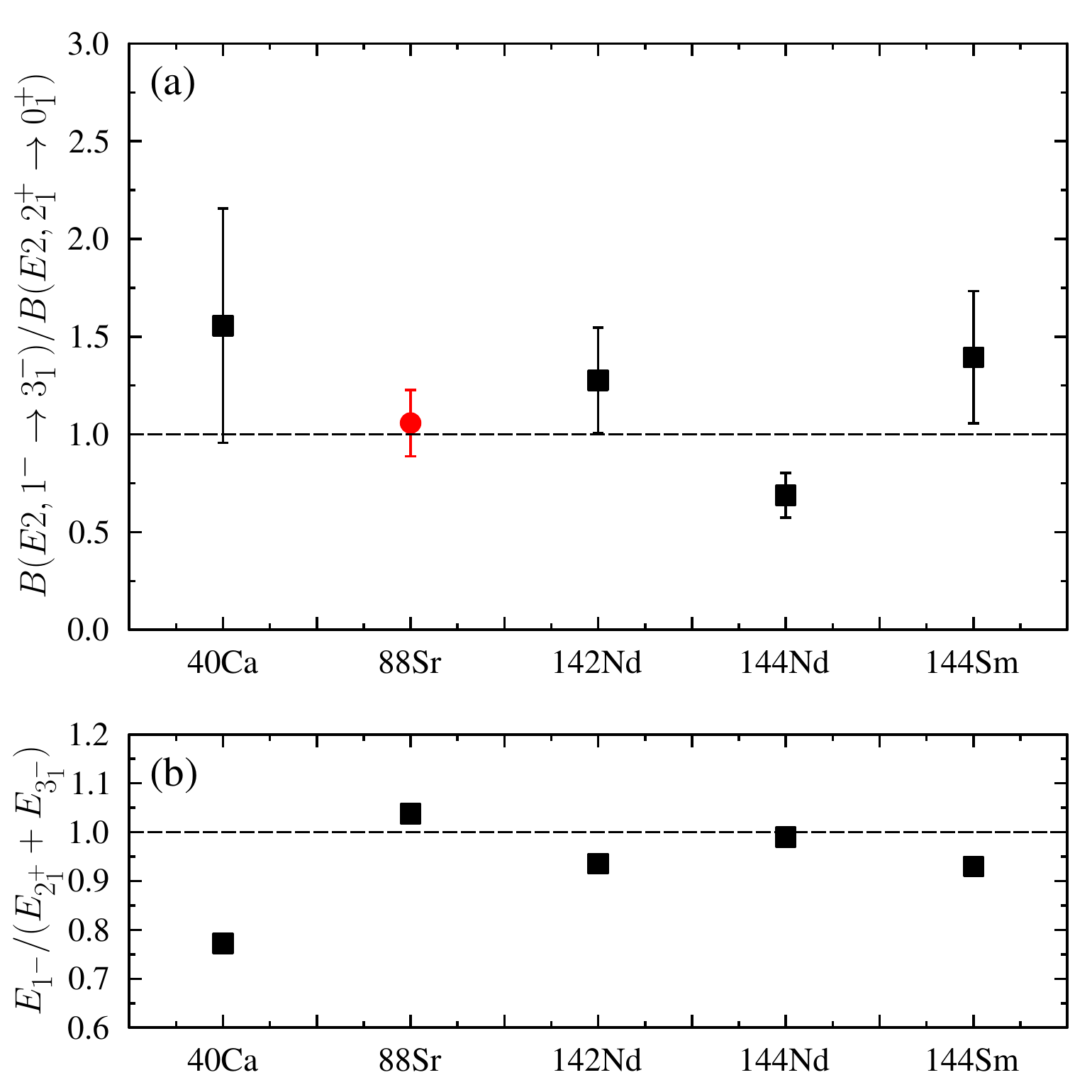}%
\caption{Comparison of (a) the experimentally determined $B(E2)$ value branching ratios for two-phonon $1^{-}_{2ph}$ states in $^{40}$Ca~\cite{dery16}, $^{142}$Nd~\cite{wilh96,wilh98}, $^{144}$Nd~\cite{sonz01}, and $^{144}$Sm~\cite{wilh96} and (b) the corresponding excitation energies. \label{fig::be2ratio}}
\end{figure}

Experimental information on the $B(E2, 1^{-}_{2ph} \to 3^{-}_{1}) / B(E2, 2^{+}_{1} \to 0^{+}_{1})$ ratio is very scarce and has been observed only for a few nuclei so far, namely $^{40}$Ca~\cite{dery16}, $^{142}$Nd~\cite{wilh96,wilh98}, $^{144}$Nd~\cite{sonz01}, and $^{144}$Sm~\cite{wilh96}.
The corresponding measured $B(E2)$ ratios are displayed in Fig.~\ref{fig::be2ratio}~a).
The dashed line indicates 1.0, which is the expected ratio in an unperturbed phonon-coupling picture. 
A very good agreement with the two-phonon interpretation of the lowest-lying $1^{-}$ states is observed in most cases while the $B(E2)$ value for $^{144}$Nd lies slightly below the theoretical prediction for a pure two-phonon state. 
Moreover, the excitation energies of all of these $1^-$ states lie 
within less than 10~\% of the sum energy of the constituent one-phonon states  
with the only exception of $^{40}$Ca where the $1^-_{2ph}$ state exhibits 
an slightly larger negative anharmonicity of -22~\%; compare with Fig.~\ref{fig::be2ratio}~b). 
This appreciable overall agreement of the excitation energies of the 
$1^-_{2ph}$ states, identified independently via their collective $E2$ decay strengths of the quadrupole-phonon annihilating transitions as quite unperturbed two-phonon structures, with the sum energy of the one-phonon states underpins that their coupling seems to be fairly harmonic. 

The fact, that the known lowest-lying $1^{-}$ states of nuclei at and near shell closures in different mass regions ranging from $A = 40$ to $A = 88$ and $A \approx 144$ show a similar behavior with respect to their decay properties, is clear evidence that two-phonon states are a general feature of atomic 
nuclei giving proof of the constituent quadrupole and octupole phonons as solid building blocks 
of multi-phonon nuclear structures.

\section{Conclusions}
\label{sec:con}

In summary, neutron-activation measurements were performed at the TRIGA Mark II reactor at Johannes Gutenberg-Universit\"at Mainz 
to investigate the structure of the best candidate for the quadrupole-octupole coupled two-phonon $1^{-}_{2ph}$ state at \qty{4742}{keV} of $^{88}$Sr.
The $\gamma$-decay branching ratios of the  $1^{-}_{2ph}$ state to the $2^{+}_{1,2}$ and $3^{-}_{1}$ states were measured with a precision of better than \qty{8}{\percent} improving the existing literature values by about one order of magnitude.
As a consequence, the $\Gamma_{0}/\Gamma = 0.811(5)$ value of the dominant ground-state decay channel is determined to \qty{0.6}{\percent} accuracy.
The direct transition from the $1^{-}_{2ph}$ to the $3^{-}_{1}$ was observed for $^{88}$Sr.
A comparison of the corresponding $B(E2, 1^{-}_{2ph} \to 3^{-}_{1})$ value to the ground-state decay strength of the $2^{+}_{1}$ state, $B(E2, 2^{+}_{1} \to 0^{+}_{1})$, directly proves the $1^{-}_{2ph}$ state as the two-phonon member of the ($2^{+} \otimes 3^{-}$) quintuplet.
This assignment is in remarkable agreement with the predictions of the EDF+three-phonon QPM calculations. 

While the $E2$ decay strength to the $3^{-}_{1}$ state is the decisive signatures of a quadrupole-octupole coupled two-phonon state, additional criteria, apart from the excitation energy of the $1^{-}_{2ph}$ state, were proposed for supporting such an identification~\cite{garr18}. 
For instance, transfer reactions were used to rule out two-phonon states, because they may be expected to possess small  single- and two-nucleon transfer cross sections (see, e.g., Ref.~\cite{jami14}). 
Hence, high-resolution transfer-reaction experiments 
may help to shed more light on the microscopic structure of  the $1^{-}_{2ph}$ state at \qty{4742}{keV} as a member of the quadrupole-octupole quintuplet of $^{88}$Sr.

Furthermore, extended systematic studies of the $B(E2, 1^{-} \to 3^{-}_{1})$ values of other $1^{-}$ states in the \qty{4}{MeV} to \qty{6}{MeV} excitation-energy region of $^{88}$Sr can elucidate whether the excellent match with the $B(E2, 2^{+}_{1} \to 0^{+}_{1})$ value is an exclusive property of the $1^{-}_{2ph}$ state at \qty{4742}{keV} or if other $1^{-}$ states could exhibit  similar, large decay strengths.
An experimental program on the $E2$ strength distribution of 
$1^- \to 3^-$ $\gamma$-ray transitions in vibrational nuclei 
is needed to clarify that question. 
Quasimonochromatic beams of photons may be utilized for this task. 
They can provide a selective probe for the study of dipole-excited states. 
Photonuclear reactions with quasi-monochromatic photon beams combined with a high-efficiency $\gamma$-ray coincidence setup have proven lately 
to be very well suited for systematic investigations of decay branchings of dipole-excited states in the order of a few percent~\cite{loeh13, loeh16, dery16, isaa19, isaa21, zilg22} and systematic measurements of the distribution 
of $B(E2; 1^- \to 3^-)$ values in selected vibrational key nuclei 
would be highly desirable.

\section*{Acknowledgments}
The authors thank the entire staff of the TRIGA reactor in Mainz for the excellent experimental conditions and support during the project.
Parts of the work were supported by the Deutsche Forschungsgemeinschaft (DFG, German Research Foundation) - Project-ID 499256822 - GRK 2891.
J.~I. and N.~P. acknowledge support by the grant "Nuclear Photonics" within the LOEWE program of the State of Hesse and within the Research Cluster ELEMENTS (Project ID 500/10.006). N.~T. acknowledges support by the contract PN 23.21.01.06 sponsored by the Romanian Ministry of Research, Innovation and Digitalization.
A.~Z. acknowledges support by the DFG (Contract No. ZI 510/10-1).
H.~L. is supported by DFG grant Le439/16. N.~T. is grateful for the kind hospitality of the University of Giessen.

\bibliography{bibtex_part1, bibtex_part2}
\bibliographystyle{apsrev4-2}

\end{document}